\documentclass[journal]{IEEEtran}

\usepackage{amsmath,amssymb,amsthm}
\usepackage{bm}
\usepackage{graphicx}
\usepackage{cite}
\usepackage{url}
\usepackage{algorithm}
\usepackage{algorithmic}
\usepackage{booktabs}
\usepackage{array,makecell}
\usepackage{xcolor}          
\newcommand{\sout}[1]{%
  \bgroup\markoverwith{\rule[0.5ex]{2pt}{0.5pt}}\ULon{#1}\egroup}


\newcommand{\Rx}{\mathbf{R}_x}
\newcommand{\diag}{\mathrm{diag}}
\newcommand{\sgn}{\mathrm{sgn}}
\newcommand{\tr}{\mathrm{tr}}

\theoremstyle{plain}

\theoremstyle{remark}

\theoremstyle{definition}

\theoremstyle{assumption}

\newcommand{\bw}{\mathbf{w}}
\newcommand{\bx}{\mathbf{x}}
\newcommand{\bo}{\mathbf{w}_o}
\newcommand{\wt}{\tilde{\mathbf{w}}}
\newcommand{\ew}{\mathbf{e}_w}
\newcommand{\ws}{\mathbf{w}_s}

\begin{document}

\title{A Double Proportionate Sparse Adaptive Filter for Impulsive Noise Environments}

\author{Mohammad Salman, \IEEEmembership{Senior Member, IEEE}, Falah Ibrahim Al Hassan, Hadi Zayyani, \IEEEmembership{Member, IEEE}, Felipe A. P. de Figueiredo, \IEEEmembership{Senior Member, IEEE} and Hasan Abu Hilal \IEEEmembership{Member, IEEE}

\thanks{M. Salman is with College of Engineering and Technology, American University of the Middle East, Egaila, 54200, Kuwait (e-mail: mohammad.salman@aum.edu.kw).}
\thanks{F. I. Al Hassan is with Ahliyya Amman University, Amman, Jordan (e-mail: f.alhassan@ammanu.edu.jo).}
\thanks{H. Zayyani and F. A. P. de Figueiredo are with the National Institute of Telecommunications (Inatel), Santa Rita do Sapucaí, Brazil. (e-mail: hadi.zayyan@posdoc.inatel.br, felipe.figueiredo@inatel.br).}
\thanks{H. Abu Hilal is with the Department of Electrical Engineering, Higher Colleges of Technology, Abu Dhabi, UAE. (e-mail: hasan.abuhilal@hct.ac.ae).}
}

\maketitle

\begin{abstract}
Sparse adaptive filters and impulsive noise robust algorithms have largely been developed along separate tracks, leaving a gap when both properties are needed simultaneously. This letter proposes the double proportionate sparse adaptive filter (DP-SAF), which closes this gap within a single $\mathcal{O}(M)$ update. Two independent diagonal gain matrices are introduced; one scales the adaptation step proportionately to coefficient magnitudes, and the other applies a magnitude-dependent zero-attraction that is strongest for inactive taps. A sign-error update provides robustness against impulsive corruptions. Both gain matrices are derived from a minimum-norm optimization framework. Simulations under a Bernoulli impulsive noise model show that DP-SAF consistently achieves a better steady-state MSD than the competing algorithms while matching or exceeding their convergence speeds.
\end{abstract}


\section{Introduction} \label{sec:intro}

\IEEEPARstart{A}{daptive} filters are widely used in system identification , channel estimation, and acoustic echo cancellation \cite{sayed2003}. The least mean square (LMS) algorithm \cite{Widrow1960} remains the most common starting point, owing to its simplicity and low complexity. It has two well-known weaknesses, however, that are hard to avoid in practice.

The first weakness is that LMS assigns the same gain to every coefficient, which is wasteful when the unknown system is sparse. PNLMS \cite{Duttweiler2000} addressed this by scaling each coefficient's step size with its own magnitude, directing more
adaptation energy toward the active taps. The proportionate idea has since been extended to affine projection structures \cite{Gay1998} and continues to evolve \cite{Luo2022, Zhuang2022, JNi2023, cao2024}. The second weakness is sensitivity to impulsive noise: a single large error sample can badly perturb the weight vector under a squared-error criterion, a problem that arises in power line communications, underwater acoustics, and wireless systems \cite{Salman25}. Replacing the raw error with its sign limits the damage any one impulse can cause \cite{Mathews87} and
underlies a family of robust algorithms. The RMN algorithm \cite{Chambers1997} blended $\ell_1$ and $\ell_2$ norms to balance robustness against efficiency. CMPN \cite{Zayyani2014} formalized this by integrating $p$-norm criteria over $p\in[1,2]$, behaving like $\ell_1$ near the origin and recovering $\ell_2$ efficiency for moderate errors. Its generalization GCMPN \cite{GCMPN2019} added
first and second-order derivative terms for a finer convergence--accuracy
tradeoff. LNAF \cite{huang2022} embedded a sliding-window Lawson cost in an affine projection framework, while LCMN \cite{Zayyani2024} introduced exponential weighting to further suppress large errors, at the cost of a slower transient.

A complementary strategy pushes inactive coefficients toward zero through zero-attraction penalties \cite{Chen2009, Salman15, Bhattacharjee2020}. RZA-LMS \cite{Chen2009} sharpened this by concentrating the penalty on near-zero coefficients while leaving large active taps largely undisturbed. VSS-RZA-LMM \cite{Wang2020} went further, pairing reweighted zero-attraction with an M-estimator
cost and an adaptive step size to simultaneously exploit sparsity, track time-varying systems, and withstand impulsive noise. Proportionate correntropy-based methods have also shown promise in non-Gaussian settings \cite{huang2024}.

What remains missing, however, is an algorithm that brings all three ingredients together in a single $\mathcal{O}(M)$ update: independent proportionate gain matrices, one steering adaptation energy toward active taps and one concentrating zero-attraction on dormant ones, combined with a sign-error update for impulsive noise robustness. This letter proposes the double proportionate sparse adaptive filter (DP-SAF) to fill that gap. Two independent diagonal gain matrices are
introduced: $\mathbf{G}_1$ scales the gradient step proportionately to coefficient magnitudes, and $\mathbf{G}_2$ applies a zero-attraction that is strongest for inactive taps and weakest for active ones. A sign-error update rounds out the design. The three mechanisms are complementary by construction and reinforce one
another without trading one property for another.

In addition, the proposed algorithm is well suited for real-time and embedded implementations. Its linear complexity and diagonal structure avoid matrix inversions and allow for efficient parallel updates, making it practical for low-latency applications. The main contributions of this letter are as follows:
\begin{itemize}
\item DP-SAF is the first algorithm to apply independent proportionate weighting to both the gradient step and the zero-attraction term in a single $\mathcal{O}(M)$ update.
\item Both gain matrices are derived from a minimum-norm optimization framework. $\mathbf{G}_1$ and $\mathbf{G}_2$ follow opposite proportionality rules by design, so they work together rather than against each other.
\item The sign-error update bounds the influence of any single corrupted sample, keeping the algorithm well-behaved under impulsive noise where squared-error methods typically diverge.
\end{itemize}

\section{Proposed Algorithm} \label{sec:prop}
\noindent In this section, we introduce the proposed DP-SAF algorithm. The proposed DP-SAF is designed to address three practical limitations simultaneously; slow convergence on sparse systems, poor robustness under impulsive noise, and the absence of an active mechanism to suppress inactive coefficients. To this end, two independent diagonal gain matrices are introduced:
\begin{align}
\nonumber 
\bw(n+1) &= \bw(n)+ \mu \textbf{G}_1(n)\,\bx(n)\,\sgn(e(n))\\
&-\rho\textbf{G}_2(n)\sgn\!\left(\bw(n)\right),
\label{eq:update}
\end{align}
\noindent where $\mu > 0$ is the step size and $\rho \geq 0$ controls the strength of the zero-attraction penalty. The vector $\mathbf{w}_s(n) \triangleq \text{sgn}(\mathbf{w}(n))$ collects the element-wise signs of the current weight vector. The two gain matrices serve distinct but complementary roles:$ \mathbf{G}_1$ gives larger adaptation steps to coefficients that are already large, speeding up convergence on the active taps, while $\mathbf{G}_2$ pulls likely-inactive coefficients efficiently toward zero. Meanwhile, replacing the raw error with its sign, $\text{sgn}(e(n))$, ensures that no single large outlier can disproportionately perturb the weight vector, which is precisely the vulnerability that makes standard squared-error methods unreliable under impulsive noise. Together, the three mechanisms reinforce one another yielding faster convergence, inactive taps are pulled faster toward zero, and robust handling of impulsive noise without any of them coming at the expense of the others. The two diagonal gain matrices are constructed as follows:

\begin{equation}
\textbf{G}_1(n) = \diag\!\left(\textbf{g}_1(n)\right),\quad \mathrm{and}
\quad \textbf{G}_2(n) = \diag\!\left(\textbf{g}_2(n)\right),
\label{eq:G}
\end{equation}
\noindent where $\textbf{g}_1(n)$ and $\textbf{g}_2(n)$ are the free parameters to be determined and given as:
\begin{align}
\nonumber
\textbf{g}_1(n) &= \left[g_{1,1}(n),\ldots,g_{1,M}(n)\right]^T,\\
\textbf{g}_2(n) &= \left[g_{2,1}(n),\ldots,g_{2,M}(n)\right]^T.
\label{eq:dof}
\end{align}

\noindent The cost function jointly penalizes the steady-state estimation error and the magnitude of the weight perturbation at each step:
\begin{equation}
J(\textbf{G}_1,\textbf{G}_2) = E\{\|\wt(n+1)\|^2\} + \gamma\,E\{\|\ew(n)\|^2\},
\label{eq:J_original}
\end{equation}
where $\gamma > 0$ is a regularization parameter that controls
the trade-off between convergence speed and update stability, and
\begin{align}
\wt(n) &= \bo - \bw(n),
\label{eq:wt}\\
\ew(n) & = \mu\textbf{G}_1\bx\,\sgn(e(n)) - \rho\textbf{G}_2\ws,
\label{eq:ew}\\
\wt(n+1) &= \wt(n) - \ew(n).
\label{eq:wt_recursion}
\end{align}


\noindent Minimizing \eqref{eq:J_original} subject to a minimum-norm constraint on $(g_{1,k}, g_{2,k})$ (full derivation in the Supplementary Material) yields the optimal per-coefficient gains:

\begin{align}
    g_{1,k}^* 
    &= \frac{(1-\lambda)\,\mu x_k \sgn(e(n))\,\eta_k}{\mu^2 x_k^2 + \rho^2},
    \label{eq:g1_opt}\\[6pt]
    g_{2,k}^* 
    &= \frac{-(1-\lambda)\,\rho\,\sgn(w_k)\,\eta_k}
            {\mu^2 x_k^2 + \rho^2},
    \label{eq:g2_opt}
\end{align}
\noindent The optimal gains in \eqref{eq:g1_opt}--\eqref{eq:g2_opt} depend on
$\eta_k = E\{\tilde{w}_k(n)\}$, which is not directly observable without
prior knowledge of the unknown system $\mathbf{w}_o$. Yet their form offers
a useful guide; since both gains scale with $\eta_k$, coefficients that are
far from their true values should be updated more aggressively than those
that are already close. To make this actionable, we use the standard sparse
adaptive filtering approximation $|\eta_k| \approx c|w_k(n)|$, which reflects
a natural property of sparse systems, i.e., active taps carry larger magnitudes
and larger errors, while inactive taps hover near zero throughout adaptation.
Replacing $\eta_k$ with $|w_k(n)|$ yields implementable gains that respect
this hierarchy. In particular, $g_{1,k}(n)$ should grow with $|w_k(n)|$ to
accelerate convergence on active taps, while remaining bounded away from zero
so that every coefficient continues to adapt. This leads to the normalized
proportionate form

\begin{equation}
g_{1,k}(n) = \frac{\max\!\left(\delta,\,|w_k(n)|\right)}
{\dfrac{1}{M}\displaystyle\sum_{j=1}^{M}
\max\!\left(\delta,\,|w_j(n)|\right)},
\label{eq:g1_pnlms}
\end{equation}

\noindent where $\delta > 0$ is a small constant that prevents the gain from becoming zero.

\noindent The second gain, $g_{2,k}(n)$, controls the zero-attraction effect. Here, the behavior is intentionally reversed. Smaller coefficients, which are more likely to be inactive, should be pulled more strongly toward zero, while larger coefficients should be affected less. This complementary design follows naturally from \eqref{eq:g1_opt} and leads to a magnitude-dependent reduction in the zero-attraction strength.

\begin{equation}
g_{2,k}(n) = \frac{1}{1 + \varepsilon\,|w_k(n)|},
\label{eq:g2_rza}
\end{equation}

\noindent where $\varepsilon > 0$ controls how sharply the attraction weakens as 
the magnitude of the coefficient increases.

\noindent The sign of $g^*_{2,k}$ in \eqref{eq:g2_opt} is governed by the
product $-\mathrm{sgn}(w_k)\,\eta_k$ and carries a physical meaning worth
noting. For an inactive tap, $w^o_k \approx 0$ and $\eta_k \approx -w_k$,
so the product is positive and $g^*_{2,k} > 0$; zero-attraction pulls
the coefficient toward zero, exactly as intended. For an active tap,
$\eta_k > 0$ and $\mathrm{sgn}(w_k)$ aligns with $w^o_k$, so the product
turns negative and the zero-attraction flips into a zero-repulsion, steering
the coefficient away from zero toward its true value, a sign flip that
is actually helpful for convergence. The practical gain \eqref{eq:g2_rza}
does not replicate this switching; instead it lets $g_{2,k}\to 0$ as
$|w_k|\to\infty$, muting the attraction on large coefficients rather than
reversing it. This is a deliberate choice: a strictly positive, monotonically
decreasing gain is far simpler to implement and sidesteps the instability
a sign-switching gain could introduce, especially under impulsive noise where
a single corrupted sample can momentarily deflect a coefficient. The key
behavior is preserved; strong attraction for near-zero inactive taps,
negligible perturbation for large active ones, and $\varepsilon$ gives
direct control over how sharply that transition occurs. Taken together,
$g_{1,k}$ accelerates convergence on active taps while $g_{2,k}$ shrinks
inactive ones, and the two work in concert rather than at cross-purposes. A complete summary of the proposed DP-SAF is given in Algorithm \ref{Alg:summary}.

\begin{algorithm}[H]
\caption{Summary of the proposed DP-SAF algorithm.}
\textbf{Parameters:} $\mu$, $\rho$, $\delta$, $\varepsilon$ \quad \\
\textbf{Initialize:} $\bw(0) = \mathbf{0}$\\
\textbf{For} $n = 0, 1, 2, \ldots$ \textbf{do:}
\begin{enumerate}
\item Compute error: $e(n) = d(n) - \bw^T(n)\bx(n)$.
\item Compute step-size gain for each $k$:
\[
    g_{1,k}(n) = \frac{\max(\delta,|w_k(n)|)}
                      {\frac{1}{M}\sum_{j=1}^M\max(\delta,|w_j(n)|)}.
\]
\item Compute zero-attraction gain for each $k$:
\[
    g_{2,k}(n) = \frac{1}{1+\varepsilon|w_k(n)|}.
\]
\item Update weights:
\begin{align}
    \nonumber \bw(n+1) &= \bw(n)
    + \mu\,\textbf{G}_1(n)\,\bx(n)\,\sgn(e(n))\\\nonumber
    &- \rho\,\textbf{G}_2(n)\,\sgn(\bw(n)).
\end{align}
\end{enumerate}
\textbf{Complexity:} $\mathcal{O}(M)$ per iteration (same as LMS).
\label{Alg:summary}
\end{algorithm}

\noindent A practical advantage of the proposed algorithm is that its gain matrices are diagonal, which eliminates the need for full matrix inversions (operations that scale at $\mathcal{O}(M^3)$ and are notoriously expensive in hardware-constrained environments. Instead, each diagonal entry is simply inverted independently, reducing the per-iteration complexity to $\mathcal{O}(M)$. Table \ref{tab:complexity} compares the per-iteration complexity of the proposed algorithm against the competing ones. The proposed approach achieves LMS-level efficiency while retaining the faster convergence. This combination makes it well suited for real-time applications where both speed and resources are at a premium.

\begin{table}[h]
\centering
\caption{Per-iteration complexity comparison.}\label{tab:complexity}
\begin{tabular}{|l|c|c|}
\hline
\textbf{Algorithm} & \textbf{Complexity} & \textbf{Transcendental ops} \\
\hline
LMS             & $\mathcal{O}(M)$ & none \\
PNLMS           & $\mathcal{O}(M)$ & none \\
RZA-LMS         & $\mathcal{O}(M)$ & none \\
GMN             & $\mathcal{O}(M)$ & 1 $\log$ \\
GCMPN           & $\mathcal{O}(M)$ & 1 $\log$ \\
LCMN            & $\mathcal{O}(M)$ & 1 $\exp$, 1 $\log$ \\
VSS-RZA-LMM     & $\mathcal{O}(M)$ & 1 $\exp$ \\
\textbf{DP-SAF} & $\mathcal{O}(M)$ & \textbf{none} \\
\hline
\end{tabular}
\end{table}

\section{Convergence Analysis}\label{sec:conv}

\noindent The analysis , in this section, relies on four standard assumptions: \textbf{A1)}
$\bx(n)$ is zero-mean, stationary with $\Rx = E\{\bx\bx^T\} \succ \mathbf{0}$;
\textbf{A2)} $v(n)$ is independent of $\bx(n)$ with impulse amplitude satisfying $|\kappa|\gg|\wt^T\bx|$; \textbf{A3)} $\bw(n)$ is independent of $\bx(n)$ (independence assumption); and \textbf{A4)} the gains are slowly varying, so
$E\{\mathbf{G}_i(n)\}\approx\bar{\mathbf{G}}_i$, $i=1,2$.

\subsection{Mean Convergence}
\noindent Subtracting $\bo$ from \eqref{eq:update} gives the weight-error recursion $\wt(n+1)=\wt(n)-\mu\mathbf{G}_1\bx\,\sgn(e) +\rho\mathbf{G}_2\ws$, where $e(n)=\wt^T\bx+v$. Under \textbf{A2}, for large $|\kappa|$:
$E_v\{\sgn(\xi+v)\}\approx(1-p)\,\sgn(\xi)$, since the symmetric impulse components cancel in expectation. Applying Price's theorem \cite{sayed2003} to the Gaussian term
$\xi=\wt^T\bx\sim\mathcal{N}(0,\sigma_a^2)$, where $\sigma_a^2=\wt^T\Rx\wt$, yields
$E\{\bx\,\sgn(\xi)\}=\sqrt{2/\pi}\,\Rx E\{\wt\}/\sigma_a$.
Taking expectations of the error recursion under \textbf{A1--A4}:

\begin{align}
E\{\wt(n+1)\}
&= \!\left(\mathbf{I} - \frac{\mu(1-p)\sqrt{2/\pi}}{\sigma_a(n)}\bar{\mathbf{G}}_1\Rx\right)\!E\{\wt(n)\}\nonumber\\
&+ \rho\bar{\mathbf{G}}_2 E\{\ws\}.
\label{eq:mean_rec}
\end{align}

\noindent The zero-attraction term $\rho\bar{\mathbf{G}}_2 E\{\ws\}$ vanishes for
inactive taps as $w_k\!\to\!0$ and is negligible for active taps
when $\rho/\mu\!\leq\!0.30$. Mean convergence is guaranteed when
the driving matrix in \eqref{eq:mean_rec} has spectral radius
less than one, giving the stability condition:

\begin{equation}
0 < \mu <
\frac{\sigma_a(n)\sqrt{2\pi}}{(1-p)\,\lambda_{\max}(\bar{\mathbf{G}}_1\Rx)}.
\label{eq:stab}
\end{equation}

\subsection{Steady-State MSD}
\noindent Expanding $\|\wt(n+1)\|^2$ from the error recursion and taking expectations under \textbf{A1--A4}: (i)~$E\{\wt^T\mathbf{G}_1\bx\,\sgn(e)\}\approx
(1-p)\sqrt{2/\pi}\,\Xi/\sigma_a$, where $\Xi\!=\!E\ \wt^T\bar{\mathbf{G}}_1\Rx\wt\}\!=\!\alpha\sigma_a^2$
with $\alpha\!\in\![\lambda_{\min}(\bar{\mathbf{G}}_1),\lambda_{\max}(\bar{\mathbf{G}}_1)]$;
(ii)~for inactive taps ($w_k^o\!=\!0$), $s_k\!=\!-\sgn(\tilde{w}_k)$, so
$E\{\wt^T\mathbf{G}_2\ws\}\approx-\Psi\leq 0$,
where $\Psi\!=\!\sum_{k\in\mathcal{I}}\bar{g}_{2,k}E\{|\tilde{w}_k|\}$;
(iii)~$E\{\|\mathbf{G}_1\bx\|^2\}\approx\tr(\bar{\mathbf{G}}_1^2\Rx)$
and $E\{\|\mathbf{G}_2\ws\|^2\}=\tr(\bar{\mathbf{G}}_2^2)$.
Setting $E\{\|\wt(n+1)\|^2\}=E\{\|\wt(n)\|^2\}$ at steady state
and solving:
\begin{equation}
\sigma_{a,\infty}=\frac{\sqrt{\pi/2}\left[\mu^2\tr(\bar{\mathbf{G}}_1^2\Rx)+\rho^2\tr(\bar{\mathbf{G}}_2^2)-2\rho\Psi_\infty\right]}{2\mu(1-p)\alpha}.
\label{eq:msd}
\end{equation}

\noindent Three observations follow from \eqref{eq:msd}. First, setting $\rho\!=\!0$ recovers the proportionate sign-LMS baseline.  Second, $\Psi_\infty\!>\!0$ for any sparse system, so moderate $\rho$ lowers the MSD floor relative to algorithms
without zero-attraction, consistent with the recommended $\rho/\mu\!\in\![0.05,0.30]$.  Third, the $(1-p)$ denominator shows graceful degradation under impulsive noise, contrasting sharply with squared-error methods where a single impulse of amplitude $\kappa$ raises the MSD by $O(\kappa^2)$.

\section{Simulation Results} \label{sec:sim}
\noindent We evaluated DP-SAF on a sparse system identification task with
impulsive measurement noise, averaging all results over 200 independent
Monte Carlo trials. The unknown system was a sparse FIR filter of length
$M = 32$ with 10\% nonzero coefficients (roughly 3 active taps), whose
positions were drawn randomly in each trial from a standard Gaussian
distribution. This setup represents acoustic echo cancellation and
underwater acoustic channels, where most taps carry negligible energy.
The input was a correlated $\mathrm{AR}(2)$ sequence with coefficients
$b_1 = 0.4$, $b_2 = -0.4$, and unit-variance white Gaussian driving noise,
chosen over white input to better reflect the spectral character of speech
and communications signals. Impulsive noise was injected via the Bernoulli
model

\begin{equation}
\eta(n) = \begin{cases} \kappa & \text{with probability } p \\ 
                        0      & \text{with probability } 1-p \end{cases}
\end{equation}

\noindent with probability $p = 0.2$ and amplitude $\kappa = 100$, scaled
by $0.3873$, giving the observed signal $y(n) = \mathbf{w}^T\mathbf{x}(n)
+ 0.3873\,\eta(n)$. With one in every five samples corrupted on average,
this is a genuinely harsh setting for any algorithm that relies on a
squared-error criterion.

\noindent DP-SAF was compared against seven algorithms. LMS served as the
plain baseline with no sparsity exploitation, no impulsive noise protection.
PNLMS represented the proportionate family, directing larger update gains
toward coefficients with larger magnitudes. RZA-LMS \cite{Chen2009} added
zero-attraction concentrated on near-zero coefficients, making it the natural
reference for isolating the contribution of DP-SAF's $\mathbf{G}_2$ component.
VSS-RZA-LMM \cite{Wang2020} is the closest competitor in design, pairing
reweighted zero-attraction with an M-estimator cost and an adaptive step size
to tackle sparsity and impulsive noise together. The remaining three 
GMN \cite{GMN2017}, GCMPN, and LCMN all replace the raw error with its
sign, making them relevant benchmarks for the robustness side of the
comparison.

\noindent The performance of the algorithms was assessed using MSD, defined as $\mathrm{MSD}(n) = 10\log_{10} \frac{\left\|\mathbf{w} - \hat{\mathbf{w}}(n)\right\|^{2}} {\left\|\mathbf{w}\right\|^{2}},$ where $\mathbf{w}$ is the true sparse channel vector and $\hat{\mathbf{w}}(n)$ is the filter estimate at iteration $n$. 

\begin{figure}[t!]
    \centering
    \includegraphics[width=0.75\linewidth]{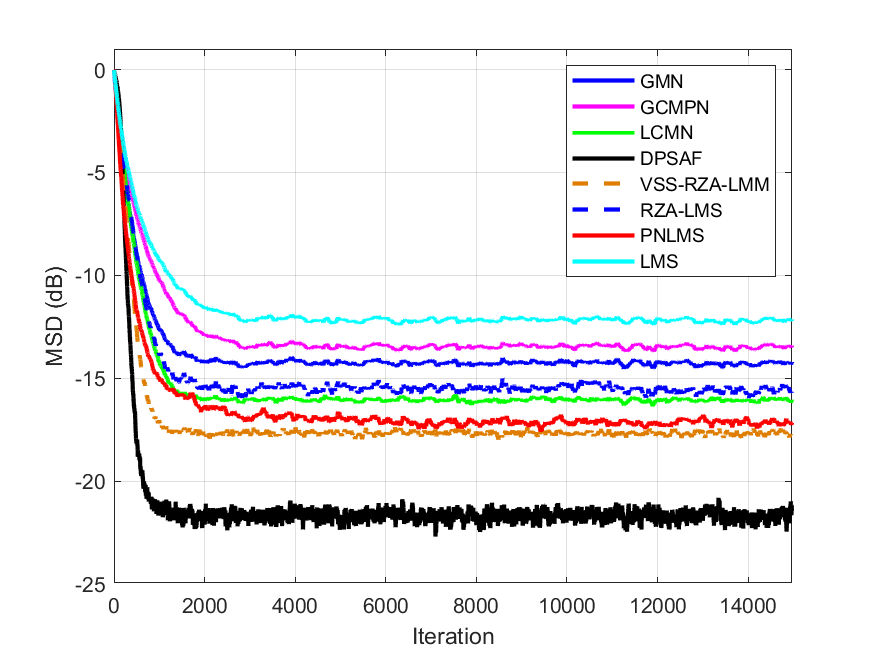}
    \vspace{-0.1cm}
    \caption{MSD learning curves for LMS, PNLMS, GMN, GCMPN, LCMN, RZA-LMS, VSS-RZA-LMM, and the proposed DPSAF under impulsive noise ($M=32$ for filter length and  $N=15000$ for iterations).}
    \label{Fig1}
\end{figure}

\begin{figure}[t!]
    \centering
    \includegraphics[width=0.75\linewidth]{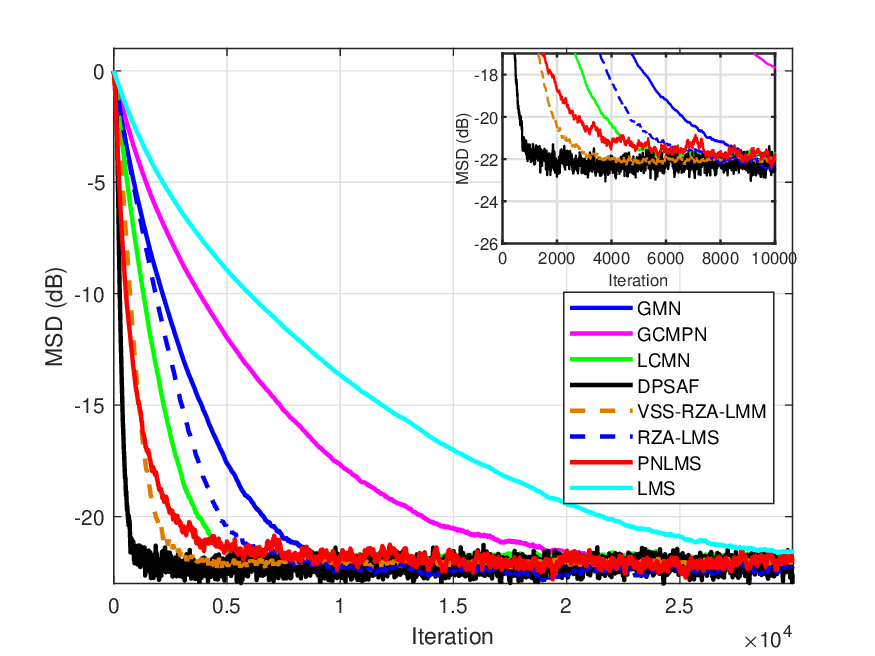}
    \vspace{-0.1cm}
    \caption{MSD learning curves for LMS, PNLMS, GMN, GCMPN, LCMN, RZA-LMS, VSS-RZA-LMM, and the proposed DPSAF under impulsive noise ($M=32$ for filter length and  $N=30000$ for iterations).}
    \label{Fig2}
\end{figure}

\noindent The purpose of the simulated experiments is to evaluate the proposed DP-SAF against the seven competing algorithms under impulsive noise conditions. All results are generated with algorithm parameters listed in Table \ref{tab:params}. The two experiments target different aspects of algorithm behavior. In Experiment 1, Fig. \ref{Fig1} shows that all algorithms reach steady state at roughly the same point, yet their final MSD floors differ considerably. The proposed DP-SAF settles at the lowest floor among all competitors, which shows that the dual-gain design genuinely improves steady-state accuracy rather than simply trading convergence speed for a lower floor. In Experiment 2, Fig. \ref{Fig2} shows that, when all algorithms are given enough iterations to converge to the same steady-state MSD, DP-SAF gets there noticeably faster than the rest. This confirms that the performance advantage observed in Experiment 1 is not purely a steady-state phenomenon, but the dual proportionate gain structure also accelerates the transient, confirming that the dual-gain design does not trade one property for the other.

\noindent Table \ref{tab:params} lists the parameter settings used for each algorithm in the simulations.

\begin{table}[h]
\centering
\caption{Algorithm Parameters for Experiments 1 and 2}
\label{tab:params}
\footnotesize
\setlength{\tabcolsep}{3pt}
\renewcommand{\arraystretch}{1.25}
\begin{tabular}{@{}lp{3.1cm}p{3.0cm}@{}}
\toprule
\textbf{Algorithm} & \textbf{Exp.\,1} & \textbf{Exp.\,2} \\
\midrule
LMS
  & $\mu=2.5\!\times\!10^{-3}$
  & $\mu=2.8\!\times\!10^{-4}$ \\[2pt]
GMN
  & $\mu=3\!\times\!10^{-3}$
  & $\mu=6\!\times\!10^{-4}$ \\[2pt]
GCMPN
  & $\mu=1.5\!\times\!10^{-3},\ \theta=5$
  & $\mu=2.5\!\times\!10^{-4},\ \theta=5$ \\[2pt]
LCMN
  & $\mu=3,\ s=5$
  & $\mu=1,\ s=5$ \\[2pt]
VSS-RZA-LMM
  & $\mu=0.02,\ \rho=5\!\times\!10^{-4}$, $\alpha=0.5,\ \eta=3,\ \varepsilon=5$
  & $\mu=0.0085,\ \rho=10^{-4}$, $\alpha=0.5,\ \eta=3,\ \varepsilon=5$ \\[2pt]
RZA-LMS
  & $\mu=0.003,\ \rho=5\!\times\!10^{-4}$, $\varepsilon=5$
  & $\mu=7\!\times\!10^{-4},\ \rho=1.5\!\times\!10^{-4}$, $\varepsilon=5$ \\[2pt]
PNLMS
  & $\mu=0.03,\ \rho=0.05$, \quad $\delta=0.01,\ \zeta=10^{-5}$
  & $\mu=0.01,\ \rho=0.01$, \quad $\delta=0.01,\ \zeta=10^{-5}$ \\[2pt]
DP-SAF
  & $\mu=0.002,\ \rho=3\!\times\!10^{-4}$, $\delta=0.1,\ \varepsilon=5$
  & $\mu=0.003,\ \rho=5\!\times\!10^{-4}$, $\delta=0.1,\ \varepsilon=5$ \\
\bottomrule
\end{tabular}
\end{table}

\section{Conclusion}\label{sec:conc}
This letter introduced DP-SAF, a sparse adaptive filter that simultaneously addresses three limitations that often co-occur in practice: slow convergence on sparse systems, poor steady-state accuracy under impulsive noise, and the absence of a mechanism to suppress inactive coefficients. The key insight is that sparsity
exploitation and zero-attraction regularization call for opposite proportionality rules, and encoding each through a separate diagonal gain matrix keeps the overall complexity at $\mathcal{O}(M)$. A sign-error update caps the influence of any single corrupted sample, providing robustness without sacrificing efficiency. Simulations
confirmed that DP-SAF achieved the lowest steady-state MSD in Experiment 1 and the fastest convergence in Experiment 2, showing that the dual-gain structure improves both accuracy and transient speed without trading one for the other. Extending DP-SAF to complex-valued systems and inputs with strong spectral coloring remains an open and promising direction.


\end{document}